\documentclass[10pt,twocolumn,aps, prb, showpacs,preprintnumbers,amsmath,amssymb]{revtex4-1}

\usepackage{graphicx}
\usepackage{dcolumn}
\usepackage{bm}
\usepackage[percent]{overpic}
\usepackage{color}

\begin{document}
\newcommand{\Arg}[1]{\mbox{Arg}\left[#1\right]}
\newcommand{\hash}{\mbox{\#\#\#\#\#\#}}
\newcommand{\bb}{\mathbf}
\newcommand{\braopket}[3]{\left \langle #1\right| \hat #2 \left|#3 \right \rangle}
\newcommand{\braket}[2]{\langle #1|#2\rangle}
\newcommand{\be}{\[}
\newcommand{\br}{\vspace{4mm}}
\newcommand{\bra}[1]{\langle #1|}
\newcommand{\braketbraket}[4]{\langle #1|#2\rangle\langle #3|#4\rangle}
\newcommand{\braop}[2]{\langle #1| \hat #2}
\newcommand{\dd}[1]{ \! \! \!  \mbox{d}#1\ }
\newcommand{\DD}[2]{\frac{\! \! \! \mbox d}{\mbox d #1}#2}
\renewcommand{\det}[1]{\mbox{det}\left(#1\right)}
\newcommand{\ee}{\]} 
\newcommand{\eg}{\textbf{\\  Example: \ \ \ }}
\newcommand{\Imag}{\mbox{Im}}
\newcommand{\ket}[1]{|#1\rangle}
\newcommand{\ketbra}[2]{|#1\rangle \langle #2|}
\newcommand{\kp}{\arccos(\frac{\omega - \epsilon}{2t})}
\newcommand{\ldos}{\mbox{L.D.O.S.}}
\renewcommand{\log}[1]{\mbox{log}\left(#1\right)}
\newcommand{\Log}{\mbox{log}}
\renewcommand{\ln}[1]{\mbox{ln}\left(#1\right)}
\newcommand{\Modsq}[1]{\left| #1\right|^2}
\newcommand{\nb}{\textbf{Note: \ \ \ }}
\newcommand{\op}[1]{\hat {#1}}
\newcommand{\opket}[2]{\hat #1 | #2 \rangle}
\newcommand{\occ}{\mbox{Occ. Num.}}
\newcommand{\Real}[1]{\mbox{Re}\left(#1\right)}
\newcommand{\so}{\Rightarrow}
\newcommand{\sol}{\textbf{Solution: \ \ \ }}
\newcommand{\thetafn}[1]{\  \! \theta \left(#1\right)}
\newcommand{\tin}{\int_{-\infty}^{+\infty}\! \! \!\!\!\!\!}
\newcommand{\Tr}[1]{\mbox{Tr}\left(#1\right)}
\newcommand{\kb}{k_B}
\newcommand{\rad}{\mbox{ rad}}

\newcommand*{\citen}[1]{%
  \begingroup
    \romannumeral-`\x 
    \setcitestyle{numbers}%
    \cite{#1}%
  \endgroup   
}

\preprint{APS/123-QED}

\title{RKKY interaction between extended magnetic defect lines in graphene}

\author{P. D. Gorman$^{(1)}$}\email{pgorman@tcd.ie}
\author{J. M. Duffy$^{(1)}$}
\author{S. R. Power$^{(2)}$}
\author{M. S. Ferreira$^{(1, 3)}$}

\affiliation{
1) School of Physics, Trinity College Dublin, Dublin 2, Ireland \\
2) Center for Nanostructured Graphene (CNG), DTU Nanotech, Department of Micro- and Nanotechnology, Technical University of Denmark, DK-2800 Kongens Lyngby, Denmark\\
3) CRANN, Trinity College Dublin, Dublin 2, Ireland 
}

\date{\today}

\begin{abstract}
Of fundamental interest in the field of spintronics is the mechanism of indirect exchange coupling between magnetic impurities embedded in metallic hosts.
A range of physical features, such as magnetotransport and overall magnetic moment formation, are predicated upon this magnetic coupling, often referred to as the Ruderman-Kittel-Kasuya-Yosida (RKKY) interaction.
Recent theoretical studies on the RKKY in graphene have been motivated by possible spintronic applications of magnetically doped graphene systems.
In this work a combination of analytic and numerical techniques are used to examine the effects of defect dimensionality on such an interaction.
We show, in a mathematically transparent manner, that moving from single magnetic impurities to extended lines of impurities effectively reduces the dimensionality of the system and increases the range of the interaction.
This has important consequences for the spintronic application of magnetically-doped and we illustrate this with a simple magnetoresistance device.

\end{abstract}

                 
\maketitle 
\section{Introduction}
\label{sec:intro}

  Interest in graphene has been spreading within the scientific community due to its potential for applications in myriad fields such as photonics, sensor technology, and spintronics. \cite{riseofgraphene,neto:graphrmp,yazyev:review}
  Graphene's weak spin-orbit and hyperfine interactions, which are the sources of spin-relaxation and decoherence in other materials, make spintronic applications particularly attractive.

  Of particular interest in the field of spintronics is the mechanism of the interaction, mediated by the conduction electrons of the host material, between localized magnetic moments embedded in nanoscale systems.
  This indirect exchange coupling (IEC) manifests as an energy difference between different alignments of the localized moments and is usually calculated within the Ruderman-Kittel-Kasuya-Yosida (RKKY) approximation.\cite{RKKYIEC,RKKY:RK,RKKY:K,RKKY:Y}
  This interaction has been extensively studied in graphene, nanoribbons, and nanotubes for a wide variety of impurities.
  The behaviour of this interaction has been found to depend on the host\cite{hwang:rkkygraphene,bunder:rkkygraphene,black-schaffer_importance_2010,klinovaja_rkky_2013}, impurity type\cite{black:graphenerkky, sherafati:graphenerkky, uchoa:rkkygraphene, sherafati:rkkygraphene2, kogan:rkkygraphene, disorderedRKKY}, and impurity configuration \cite{stephenreview, dugaev:rkkygraphene, saremi:graphenerkky, brey:graphenerkky}.
  Since a range of effects are predicated on exchange interactions, methods of modifying the interaction via strain, edges, magnetic fields, doping, and lattice defects have also been studied.\cite{ourpaper,ourpaper2,Vozmediano:2005, duffy_variable_2014}

  An important aspect of the RKKY interaction is the rate at which it decays as a function of the separation, $D$, between magnetic impurities.
  In undoped graphene a decay rate of $D^{-3}$ for substitutional, top-adsorbed, and bridge-adsorbed impurities is found, while a much faster decay rate of $D^{-7}$ is found for center-adsorbed impurities.\cite{saremi:graphenerkky,sherafati:graphenerkky,uchoa:rkkygraphene,ourpaper2}
  This decay rate is faster than the $D^{-2}$ decay expected for conventional two-dimensional materials and arises from the vanishing density of states at the Fermi energy in graphene. \cite{me:grapheneGF}
  This fast decay rate results in the interaction being very short ranged and any method of amplifying the coupling to extend its range could prove useful for both the experimental detection of the RKKY interaction and future spintronic applications. 

  The sign of the coupling, which determines the ferromagnetic (FM) or antiferromagnetic (AFM) alignment of the moments, should oscillate as a function of their separation, but in graphene this is masked by the coincidence of the Fermi surface and Brillouin zone. 
  This causes the sign of the coupling, within the RKKY interaction, to depend only on whether the two moments occupy the same or opposite sublattices. 
  When graphene is doped or gated, the Fermi surface no longer coincides with the Brillouin zone, so sign-changing oscillations are recovered and the interaction is found to decay as $D^{-2}$.\cite{me:grapheneGF}

  \begin{figure}
  \includegraphics[width=0.45\textwidth]{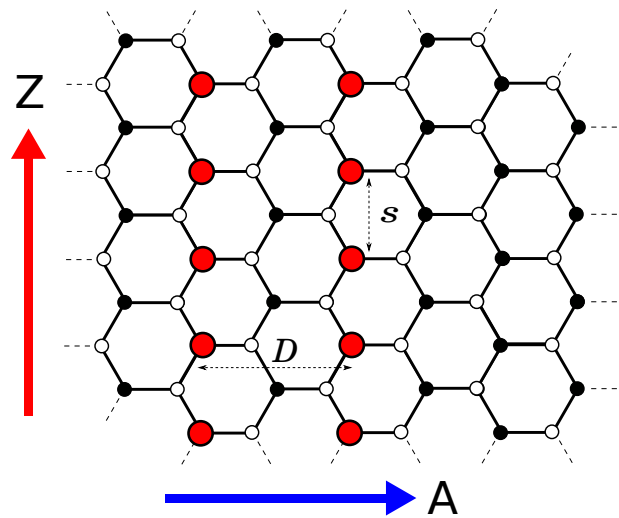}
  \caption{
  Schematic representation of the graphene lattice showing the armchair (A) and zigzag (Z) directions.
  The filled and hollow symbols ($\bullet$,$\circ$) represent sites on different sublattices.
  The red circles represent two parallel lines of magnetic impurities separated by a distance $D=1$ ( in units of 3$a$, where $a=2.64$\AA ) in the armchair direction, with the zigzag separation between moments set at $s=1$.
  }
  \label{fig:line_schematic}
  \end{figure}

  There is currently a large and growing interest in nanopatterning graphene.
  Atomically precise tailoring of 2D materials, the addition of absorbants or vacancies, allows for a complex manipulation of the electronic properties of the material.
  Recent studies have examined how larger structures of impurities or vacancies can be used to modify the electronic properties of graphene allotropes.
  One such study looks at how the controlled growth of a line of defects in graphene can be used for gate-tunable valley filtering\cite{louie_lines_2014}, while another study looks at how TM-nanowires results in long-range magnetic order and magnetic anisotropy in graphyne.\cite{he_magnetic_2014, louie_lines_2014}
  
  The possibility and ease of depositing a line of magnetic impurities in graphene has increased with the recent work of 
  Duesberg et al.\cite{winters_production_2012,hallam_field_2014}
  In these works the electronic structure of graphene is altered via folding - sometimes called graphene origami.
  This folding is achieved by depositing the graphene over a corrugated surface which, in conjunction with an applied magnetic field, has recently been shown to open a bandgap in graphene.\cite{costa_origami-based_2013}
  The ridges created during this process act as traps for magnetic impurities and open up the possibility of depositing impurities in straight parallel lines.
  Alternatively, increased reactivity near grain boundaries\cite{Salehi-Khojin2012} or simple kinks\cite{Rasmussen2013} may also lead to the formation of impurity lines. 
  A sublattice preference of the impurities along zigzag lines, as shown in Fig. \ref{fig:line_schematic}, is expected due to similar behaviour predicted for zigzag edged nanoribbons.\cite{me:impseg}
  Although the RKKY interaction in graphene has been intensively studied, one aspect that has yet to be examined is how lines of impurities, which change the dimensionality of the interaction, affect the coupling.

  With the motivation of understanding the magnetic interaction strength, and providing some theoretical foundation, we will examine the role of dimensionality in the RKKY interaction.
  In this paper we will apply analytical and numerical techniques to a system containing two parallel lines of magnetic impurities (Fig. \ref{fig:line_schematic}), and present a method for calculating the coupling between lines of impurities embedded or adsorbed in a host material. 
  Some analogues can be made between our setup and two ferromagnetic monolayers embedded in a nonmagnetic metal\cite{RKKY:Bruno1,RKKY:Bruno2}, where the Ruderman-Kittel theory can be used to  derive a $D^{-2}$ decay rate for the coupling between the ferromagnetic layers\cite{Castro:Coupling}.
  We start by introducing the general formalism used to calculate the magnetic coupling, which is written entirely in terms of the real-space single-particle Green's functions (GFs) of the host graphene sheet.
  We show in a mathematically transparent manner how the standard coupling equation may be modified to calculate the RKKY interaction between two infinite lines of impurities.
  We find that the interaction of finite lines of impurities quickly tends towards the interaction of infinite lines, and that the reduced dimensionality of the system leads to smaller decay rates and longer ranged interactions.
  These predictions are confirmed using fully numerical calculations.
  Finally, we examine the magnetoresistance response of a simple device based on the setup discussed in this paper and which may motivate further studies of graphene spintronics based on magnetic defect lines.
  In this work we only consider a graphene host, but the method is easily generalized to other two dimensional materials.


\section{Methods}
\label{sec:meth}
  The indirect exchange coupling is defined as the change in energy between impurities in FM and AFM alignments.
  This can calculated using the Lloyd Formula
  \begin{equation}
    \Delta E (E_F) = \frac{1}{\pi} \text{Im} \int dE \, f(E) \ln { \det {\hat{I} - \hat{G}\hat{V}_s}},
  \end{equation}
  where $f(E)$ is the Fermi function, $\hat{V}_s$ is the spin perturbation matrix, and $\hat{G}$ represents the GF of the system containing impurities.
  An Anderson-like Hamiltonian\cite{anderson_localized_1961} is used to describe the electronic properties of the system, and calculate the GFs.
  The impurities are introduced to the pristine system via Dyson's equation
    \begin{equation}
   \hat{G}=(\hat{I}-\hat{g} \hat{V}_I)^{-1} \hat{g},
   \label{eq:Dyson}
  \end{equation}
  where the characteristics of the impurities are contained within the perturbation matrix $\hat{V}_I$. $\hat{g}$ is the pristine graphene GF calculated from a nearest-neighbour tight-binding approximation Hamiltonian.

  To calculate the coupling between two lines of impurities of length $N$, one introduces a $2N \times 2N$ perturbation matrix $\hat{V}_I$.
  The matrix inversion present in the coupling calculation makes this computationally cumbersome as $N$ becomes large,
  however the following method allows for the rapid calculation of this coupling as well as an analytic form.
  
  The coupling calculation can be greatly simplified by keeping the GF partially in reciprocal space when introducing the perturbation. 
  We will introduce our perturbation as a line of impurities along the y (zigzag) direction, though a similar procedure may be used for impurities in x (armchair) direction (Fig.~\ref{fig:line_schematic}).
  When dealing with lines of impurities in the $y$ direction, we make the projection $ \bra{x,k_y,\alpha} \hat{g} \ket{x',k_y,\beta} =  g^{\alpha \beta}_{x x'} (k_y)$, where $\hat{g}$ is the GF of the pristine system, and $\alpha$ and $\beta$ are sublattice labels.
  This can be written as 
  \begin{equation}
    g^{\alpha \beta}_{x x'} (k_y) = \frac{1}{N_x} \sum_{k_x} \frac{ N^{\alpha \beta}(E,k_y) e^{i k_x (x'-x)} }{E^2 - |f(k_x,k_y)|^2},
  \end{equation}
  where $N^{\alpha\beta}$ is a sublattice dependent term,
  \begin{equation}
   N^{\alpha\beta} =
   \left\{
    \begin{array}{ll}
      E & \alpha= \beta \\
      f(k_x, k_y)& \alpha \neq \beta \\
    \end{array} 
    \right.
  \end{equation}
  and $f(k_x,k_y)$ is the dispersion relation.,
  \begin{equation}
   f(k_x,k_y) = t + 2 t \cos(k_x)e^{ik_y}
  \end{equation}
  The sum over $k_x$ may be converted into an integral over the $k_x$ direction in the Brillouin Zone, which can then be performed by contour integration.
  This results in an analytic form for the GF,
  \begin{equation}
    g^{\alpha \beta}_{D} (k_y) = \frac{i}{4 t^2} \frac{ N^{\alpha \beta}(E,k_y) e^{i q(ak_y) D} }{\cos\left({\frac{a k_y}{2}}\right) \sin(q(a k_y))},
  \end{equation}
  with
  \begin{equation}
    q(k_y) = {\pm} \cos^{-1} \left[ \frac{E^2 - t^2 - 4 t^2 \cos^2 \left({\frac{a k_y}{2}}\right)}{4 t^2 \cos\left({ \frac{a k_y}{2}}\right)} \right],
  \end{equation}
   where $t$ is the hopping integral and $a$ is the graphene lattice parameter.
  
  Considering Dyson's equation, we may greatly reduce our analytical workload by writing our GFs as Bloch matrices $\hat{g}_{AA} = \sum_{k_y} \ket{A,k_y} g_{A A} (k_y) \bra{A,k_y}$, so that
  \begin{equation}
    \hat{I} - \hat{g} \hat{V} = \hat{I} - 
    \begin{pmatrix}
      \hat{g}_{AA} & \hat{g}_{AB} & \cdots \\
      \hat{g}_{BA} & \hat{g}_{BB} & \\
      \vdots &  & \ddots
    \end{pmatrix}
    \begin{pmatrix}
      \tau \hat{I} & 0 & \cdots \\
      0 & \tau \hat{I} &  \\
      \vdots &  & \mathbf{0}
    \end{pmatrix}
  \end{equation}
  where our perturbation, with change in onsite energy $\tau$, has been trivially transformed to the same basis by a Fourier Transformation
  \begin{equation}
  \begin{split}
    \hat{V}_I &= \sum_a     \tau \ket{A, s}   \bra{A, s}   + \tau \ket{B, s}   \bra{B, s}, \\
            &= \frac{1}{s}\sum_{k_y} \tau \ket{A,k_y} \bra{A,k_y} + \tau \ket{B,k_y} \bra{B,k_y},
  \end{split}
  \end{equation}
  where $s$ is the integer separation between impurities along the zigzag direction as shown in Fig. \ref{fig:line_schematic}.
  Since all but the first two columns are zero we can find the desired elements by just considering the $2\times2$ matrix multiplication. 
  Similar identities facilitate the inversion, and our new GF, $\hat{G}$,  can be calculated in a relatively transparent manner
  
  The spin perturbation is introduced as
  \begin{equation}
   \hat{V}_s = \sum_{k_y}  \ket{B,k_y}\hat{V}_{\theta}\bra{B,k_y},
  \end{equation}
  where $\hat{V}_{\theta}$ is a 2D spin perturbation matrix that rotates the spins on line $B$ through an angle of $\theta=\pi$.
  Calculating $\hat{I} - \hat{g} \hat{V}$ for the spin perturbation is more involved, but the basic method is the same.
  We can now write
  \begin{equation}
  \begin{split}
    \det{ \hat{I} - \hat{G}\hat{V}_s } 
    &= \det{ (\hat{I} -2 V_{ex}\hat{g}_{BB}^{\uparrow})( \hat{I} + 2 V_{ex} \hat{g}_{BB}^{\downarrow}) },\\
    &= \prod_{k_y} [ (1 - 2 V_{ex}g_{BB}^{\uparrow}(k_y))( 1 + 2 V_{ex} g_{BB}^{\downarrow}(k_y)) ],
  \end{split}
  \end{equation}
  where all of terms are diagonal Bloch matrices in $k_y$ space, which allows us to write the determinant as a product.

  Now, we again use Dyson's equation along with the properties of logarithms to get a convenient form for the coupling per atom
  \begin{equation}
  \begin{split}
    \mathcal{J}_{AB} &= - \frac{s}{N_y \pi} \text{Im} \int_{-\infty}^{E_F} dE \\
    &\times \sum_{k_y} \ln {1 + 4 V_{ex}^2 g^{\uparrow}_{BA}(E,k_y) g^{\downarrow}_{AB}(E,k_y)}.
  \end{split}  
  \label{eq:JAB}
  \end{equation}
  The sum may be converted to an integral over the Brillouin Zone in the $k_y$ direction. 
  We may also write $k_Z = \frac{a}{2} k_y$, exploit the symmetries of the GF, and use the analyticity of the integrand in the upper half plane to change the integration to the imaginary axis.
  This gives us a form of the coupling that is suitable for evaluation using numerical methods. 
  \begin{equation}
  \begin{split}
    \mathcal{J}_{AB} &= \frac{2}{\pi^2} \int_{\eta}^{\infty} dy \, \int_{0}^{\frac{\pi}{2}} dk_Z \\
    &\times \ln{|1 + 4 V_{ex}^2 g^{\uparrow}_{BA}(E_F+iy,k_Z) g^{\downarrow}_{AB}(E_f+iy,k_Z)|}.
  \end{split}  
  \label{eq:JAB2}
  \end{equation}
  
  To obtain a useful analytic form of the coupling we make several approximations which are similar in form to the RKKY approximation.  
  We first expand the log in Eq.~\ref{eq:JAB} to first order by assuming a small exchange coupling, $V_{ex}$.
  We then make the approximation $g_{AB}^{\downarrow} = g_{BA}^{\uparrow} = g^{\alpha \beta}_{D} $ so that the coupling, which we now write as $\mathcal{J}_{D}$ to distinguish it from the numerical form above, is now given by
  \begin{equation}
   \mathcal{J}_{D} \approx -\frac{4 V_{ex}^2}{\pi^2} \text{Im} \int_{-\infty}^{E_F} dE \, \int_{-\frac{\pi}{2}}^{\frac{\pi}{2}} dk_Z \, \left[{g^{\alpha \beta}_{D}(E,k_Z)}\right]^2.
  \end{equation}
  At large separations the integrand oscillates quickly and only select points contribute to the integral over $k_Z$.
  This allows us to use the stationary phase approximation (SPA) to write the integral in a fully analytic form.
  The SPA has previously been used\cite{me:grapheneGF} to write the pristine Green's Functions of graphene in an analytic form.
  This allows one to calculate the coupling between two impurities.
  Here we use the SPA to calculate the integral over Green's Functions that already contain impurities,
  \begin{equation}
  \int_{-\frac{\pi}{2}}^{\frac{\pi}{2}} dk_Z \, \left[{g^{\alpha \beta}_{D} (E,k_Z)}\right]^2 = \frac{A(E)e^{i2\mathcal{Q}(E)D}}{\sqrt{D}},
  \end{equation}
  where $A(E)$ captures the specific sublattice dependencies, and non-contributing terms have been ignored.
  The symmetry in the zigzag direction allows us to use the SPA in this manner.
  
  We can now write our coupling equation as
  \begin{equation}
   \mathcal{J}_{D} = -\frac{4V_{ex}^{2}}{\pi} \text{Im} \int_{-\infty}^{E_F}  dE \ \frac{A(E)e^{i2\mathcal{Q}(E)D}}{\sqrt{D}} .
   \label{eq:Jspa}
  \end{equation}
  The integration procedure is now identical to that for single impurities in graphene.\cite{me:grapheneGF}
  The functions $A(E)$ and $Q(E)$ are expanded around $E_F$ and the integral can be reduced to a sum over Matsubara frequencies, which in the low temperature limit gives
  \begin{equation}
   \mathcal{J}_{D} = V_{ex}^{2} \text{Im}
   \sum_{\ell} 
   \frac
   {A^{(l)} e^{i 2 \mathcal{Q}(E) |D|} (-1)^\ell }
   {(2i\mathcal{Q}^{(1)})^{\ell+1}D^{\ell+\frac{3}{2}}}.
   \label{eq:JDA}
  \end{equation}
  The decay rate of the coupling is now determined by the first non-vanishing $A^{(\ell)}$, where this notation is defined as the $\ell^{th}$ derivative of $A$ with respect to energy evaluated at the Fermi energy.
  
%
%
  
\section{Results}
\label{sec:res}
Unless otherwise specified the results that follow refer to the case $s=1$, i.e. that the impurities are separated by the minimum distance in the zigzag direction.

\begin{figure}[h]
\includegraphics[width=0.45\textwidth]{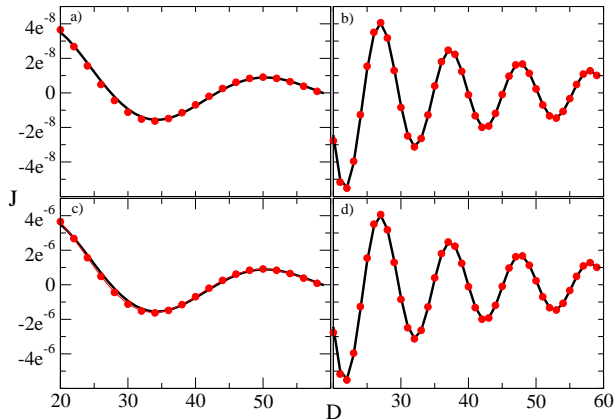}
\caption{A comparison of the coupling, $\mathcal{J}_{AB}$ (lines), and the SPA of coupling, $\mathcal{J}_{D}$ (dots), for different exchange splittings, $V_{ex}$, and at different Fermi energies, $E_F$. a) $V_{ex}=0.01$, $E_F=0.1$. b) $V_{ex}=0.01$, $E_F=0.3$. c) $V_{ex}=0.1$, $E_F=0.1$. d) $V_{ex}=0.1$, $E_F=0.3$.}
\label{fig:comparison}
\end{figure}
The stationary phase approximation (SPA) used in the derivation of Eq.\ref{eq:Jspa} relies on the assumption of large separations between the lines of impurities, so it is important to check the accuracy of this approximation at reasonable separations.
The SPA coupling, $\mathcal{J}_{D}$, was compared against the full numerical calculation of the coupling, $\mathcal{J}_{AB}$, for a variety of Fermi energies and exchange splittings.
A representative sample of these checks are shown in Fig.~\ref{fig:comparison}, where the agreement is seen to be very good even for reasonably small separations, and increases in accuracy as separation increases.
The approximation is therefore concluded to be sufficiently accurate for our uses here.

\begin{figure}
\includegraphics[width=0.45\textwidth]{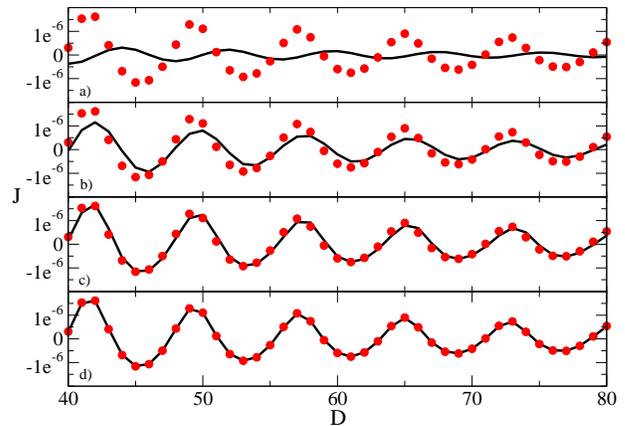}
\caption{
A plot of the numerical evaluation of the coupling, $J$, against separation, $D$, for a pair of finite lines of length a) $N=1$, v) $N=10$, c) $N=50$, d) $N=500$.
The numerical evaluation of the coupling for a pair of infinite lines (Eq.~\ref{eq:JAB2}) is represented in each subplot by the red dots.
}
\label{fig:coupling}
\end{figure}

Fig.~\ref{fig:coupling} plots the coupling between several pairs of line segments, of length $N$, against an infinite pair of lines.
As we increase the length of the line segments the magnitude and phase of the coupling are seen to approach that of the infinite line segment (red dots).
At a length of $N=500$ (Fig.~\ref{fig:coupling}d), the coupling can be seen to converge to that of an infinite line of impurities.
For shorter lines, $\sim N=50$ (Fig.~\ref{fig:coupling}c) the coupling closely resembles that between the infinite lines, with only slight discrepancies in magnitude and phase.
These discrepancies increase as the line is shortened to $\sim N=10$ (Fig.~\ref{fig:coupling}b), however even the coupling between line segments as short as $N=10$ are seen to behave much closer to the infinite case than to the case of individual impurities (Fig.~\ref{fig:coupling}a).
Thus Eq.\ref{eq:JDA} is a useful tool for quickly calculating the coupling between finite lines of impurities since it is clear that as the length, $N$, of the lines increase the interaction converges to that of the infinite case.

An important feature of the coupling is its overall rate of decay with separation, $D^{-\alpha}$, which is determined from the analytic form of the coupling in Eq.~\ref{eq:JDA}.
Here the coupling is represented as a sum of terms over $\ell$, with decay rates that increase as $D^{-\ell -\frac{3}{2}}$.
The decay rate is therefore determined by the first non-zero term in the series, as subsequent terms decay much more rapidly.
It is the quantity $\mathcal{A}^{\ell}(E)$ that determines whether or not the whole term will vanish.
For the case of lines of impurities occupying the same sublattice, where the interaction is ferromagnetic, we have
\begin{equation}
 \mathcal{A}(E) =  -\sqrt{{\frac{\pm i \pi}{(E^2 + 3t^2)}}}
 \frac{ E^2  }{2 (E^2 t^2-E^4)^{\frac{3}{4}}}.
\end{equation}
It is clear from examination that $A^{(0)}$ is non-zero whenever E is non-zero, and thus away from the Fermi energy we would expect a decay rate that goes as $\sim D^{-\frac{3}{2}}$.

\begin{figure}
\includegraphics[width=0.45\textwidth]{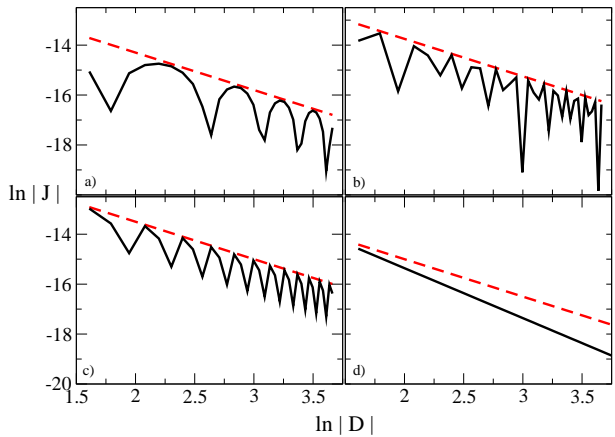}
\caption{
A log-log plot of the numerical evaluation of the coupling (J) between two lines of impurities as a function of of their separation (D) for a) $E_F=0.1$, b) $E_F=0.3$, c) $E_F=0.5$, d) $E_F=0$.
In each case the log of the coupling is represented by the solid black line and the dashed grey line represents a decay rate of $1.5$.
}
\label{fig:coupling2}
\end{figure}

In Fig.~\ref{fig:coupling2} we compare the coupling with the expected value of its decay rate at several Fermi energies.
We find that for values of the Fermi energy not equal to zero $(E_F\ne0)$ that our theoretical predictions hold.
At precisely $E_F=0$, corresponding to an undoped graphene sheet, the analytical approximation no longer holds and a faster decay rate of $\sim D^{-2}$ is found.
This difference of behaviour at the Dirac point is a hallmark of such interactions in graphene and is also seen for the single impurity case.
However, both of the linear defect results correspond to an increased interaction range compared to single site impurities, where we would expect to find decay rates of $D^{-2}$ and $D^{-3}$ for doped and undoped graphene respectively\cite{me:grapheneGF}.

The form of $\mathcal{A}(E)$ was also examined for the the case of opposite sublattice occupation, where the interaction is antiferromagnetic, as well as the case where one line is offset by half a lattice spacing.
These cases produced similar results to the initial case examined above in both coupling decay rate and the convergence of finite lines to the infinite case.
For the cases where the impurities are spaced in the zigzag direction by a larger length $s>1$, we again find that the coupling between lines quickly converges to that of the infinite case as $N$ is increased.

\section{Magnetoresistance}
RKKY interactions in multilayered systems play a key role in the Giant Magnetoresistance (GMR) effect.
Up- and down-spin electrons have different transmission probabilities through a magnetic layer of a certain orientation.
The most energetically favourable configuration of layer orientations is determined by RKKY interactions and will have a certain total transmission.
Aligning all the layers with an applied magnetic field can lead to markedly different transmission probabilities and a resulting change in the resistance of the system\cite{Fert:GMR, Grunberg:GMR}.
Previous studies have considered the possibility of magnetoresistance devices based on graphene systems\cite{Kim2008, Munoz-Rojas2009, Lu2011}, but it is worth examining whether a magnetoresistance signal emerges using devices based on the system we have considered in this work.
\begin{figure}
\includegraphics[width=0.45\textwidth]{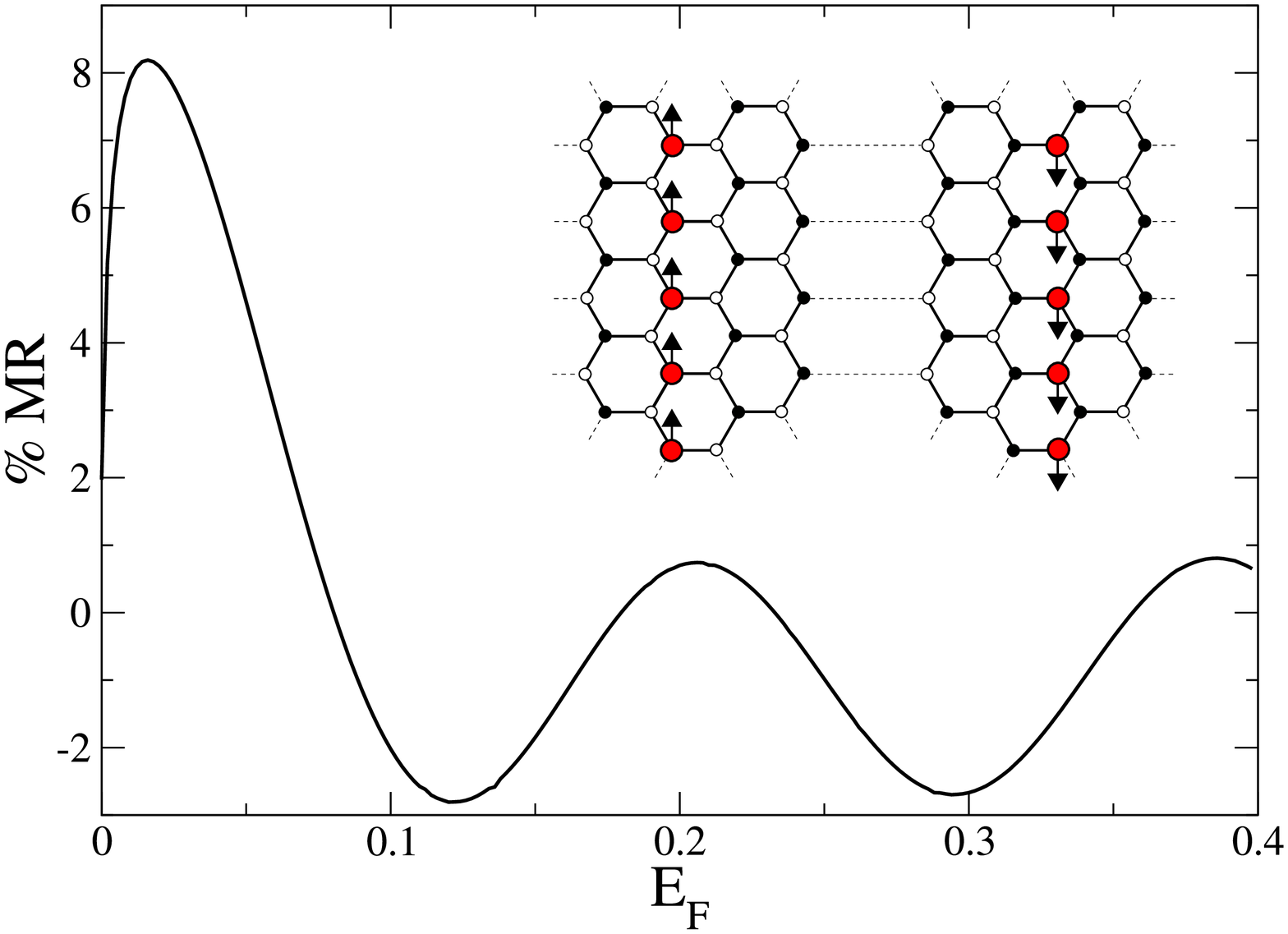}
\caption{
Magnetoresistance percentage as a function of Fermi energy for a simple device based on magnetic lines in a graphene sheet. The inset shows the initial AFM orientation of two lines of moments on opposite sublattices with $D=9$.
}
\label{fig:mr}
\end{figure}
We consider two lines of magnetic impurities, one on each sublattice, as shown in the inset of Fig. \ref{fig:mr}.
The total conductance, $\Gamma_{AFM/FM}$, through such a system is calculated as the sum of the conductances of up-spin and down-spin electrons.
These are calculated using the standard Landauer-Buttiker approach with recursively calculated Green's functions and spin-dependent potentials at each of the magnetic impurity sites.
We assume that these lines initially have an antiferromagnetic relative orientation\cite{low_energies_note} and the magnetoresistance is calculated from the relative change in the resistance when a small magnetic field is applied to force a ferromagnetic alignment
\begin{equation}
 MR = \frac{\Gamma_{FM}^{-1} - \Gamma_{AFM}^{-1}}{\Gamma_{AFM}^{-1}} \,.
\end{equation}
The magnetoresistance, expressed as a percentage, is plotted in the main panel of Fig. \ref{fig:mr}.
Although the magnitude of the magnetoresistance is quite small, it should be noted that this calculation only considers two impurity lines, whereas a more realistic device could contain many more lines and this would increase the magnetoresistance significantly.
Nonetheless, this simple example demonstrates the potential for magnetic impurity lines to play a part in graphene-based spintronics and motivates further studies of such systems and the RKKY interactions on which magnetoresistance effects are predicated.

\section{Conclusions}
\label{sec:conc}
In this work we have derived an analytic expression for the RKKY interaction in graphene between two lines of magnetic impurities separated along the armchair direction.
We have shown that the coupling between lines of impurities quickly converges to the coupling of the infinite case as their length increase.
The analytic form is therefore a useful way to approximate the interaction between line segments.
We also use this analytic form for predict the rate of decay of this interaction, away from $E_F=0$, which is slower than the rate of decay for two impurity interactions.
Furthermore we have shown that it is the reduced dimensionality of the system that increases the range of the RKKY interaction.
This increased range may ease the detection of the RKKY which is notoriously hard to examine experimentally.

Since a whole range of physical features, such as magnetotransport and overall magnetic moment formation, are predicated upon the magnetic coupling, it is hoped that this work may lead to interesting spintronic applications.


\begin{acknowledgments}
The authors acknowledge financial support received from the Programme for Research in Third-Level Institutions PRTLI5 Ireland, the Irish Research Council for Science, Engineering and Technology under the EMBARK initiative and from Science Foundation Ireland under Grant No. SFI 11/RFP.1/MTR/3083. 
The Center for Nanostructured Graphene (CNG) is sponsored by the Danish National Research Foundation, Project No. DNRF58.
\end{acknowledgments}


\begin{thebibliography}{44}
\bibitem [{low()}]{low_energies_note}%
  \BibitemOpen
  \href@noop {} {}\bibinfo {note} {This will be the case for $E_F=0$, and for
  low doping levels. As the Fermi energy is increased there will be energy
  ranges where a FM alignment is preferred and the magnetoresitance is
  zero.}\BibitemShut {Stop}%
\end{thebibliography}
%
\end{document}